\documentclass[a4paper,fleqn,usenatbib]{mnras}
\usepackage{newtxtext,newtxmath}
\usepackage{siunitx}
\usepackage[T1]{fontenc}
\usepackage{ae,aecompl}
\usepackage{graphicx}	
\usepackage{amsmath}	
\usepackage{amssymb}	

\title[Temporal variations in scattering and DM in the Crab Pulsar]{Temporal variations in scattering and dispersion measure in the Crab Pulsar and their effect on timing precision}

\author[McKee et al.]{
J.\,W.\,McKee$^{1,2}$\thanks{E-mail: jmckee@mpifr-bonn.mpg.de},
A.\,G.\,Lyne$^{1}$,
B.\,W.\,Stappers$^{1}$,
C.\,G.\,Bassa$^{3}$,
and C.\,A.\,Jordan$^{1}$
\\
$^{1}$Jodrell Bank Centre for Astrophysics, School of Physics and Astronomy, The University of Manchester, Manchester M13 9PL, UK\\
$^{2}$Max-Planck-Institut f{\"u}r Radioastronomie, Auf dem H{\"u}gel 69, 53121, Bonn, Germany\\
$^{3}$ASTRON, the Netherlands Institute for Radio Astronomy, Postbus 2, 7990 AA, Dwingeloo, The Netherlands\\
}

\date{Accepted XXX. Received YYY; in original form ZZZ}

\pubyear{2018}

\begin{document}
\label{firstpage}
\pagerange{\pageref{firstpage}--\pageref{lastpage}}
\maketitle

\begin{abstract}
We have measured variations in scattering time scales in
the Crab Pulsar over a 30-year period, using observations made at 610\,MHz with the 42-ft telescope at Jodrell Bank Observatory. Over more recent years, where regular 
Lovell Telescope observations at frequencies around 1400\,MHz were available, we have also determined the dispersion measure variations, after disentangling the scattering delay from the dispersive delay. We demonstrate a relationship between scattering and dispersion measure variations, with a correlation coefficient of $0.56\pm0.01$. 
The short time scales over which these quantities vary, the size of the variations, and the close correlation between scattering and dispersion measure all suggest that the effects are due to discrete structures within the Crab Nebula, with size scales of $\sim\!6$\,AU (corresponding to an angular size of $\sim\!2$\,mas at an assumed distance of 2200\,pc). 
We mitigate the effects of scattering on the observed pulse shape by using the measured scattering information to modify the template used for generating the pulse arrival times, thus improving the precision to which the pulsar can be timed.
We test this on timing data taken during periods of high scattering, 
and obtain a factor of two improvement in the root mean square of the timing residuals.
\end{abstract}



\section{Introduction}
The Crab Pulsar (PSR\,B0531$+$21, also known by the J2000 name PSR\,J0534+2200) was discovered in 1968 at the centre of the Crab Nebula, the remnant of its progenitor supernova SN\,1054 \citep{sr68}. As the Crab is very young, the surrounding region is energised by the pulsar spin-down energy, forming a pulsar wind nebula \citep{pac67}. The nebula contains density ripples (some of which are seen as wisps in optical observations), clouds, and irregular regions of turbulence, which reveal a dynamic internal structure (\citealp{ceh07}, \citealp{hes08}).  
As the nebula contains a large number of free electrons, radio pulses propagating through it are scattered and dispersed in a manner similar to that caused by the interstellar medium (ISM). 
Due to a combination of effects including the pulsar proper motion, the Earth's orbit, and the dynamics of the nebula, observed variations in electron number density are larger and the time scale over which they occur are shorter, compared to typical pulsars (e.g. \citealp{rr71}). This causes variations in dispersion measure and changes in the observed radio pulse shape of the Crab Pulsar to be more dramatic than those seen in typical pulsars (e.g. \citealp{lpg01}, \citealp{glj11}, \citealp{kcs+13}).
These effects are useful in understanding the influence of scattering on pulsar observations and the nature of the Crab Nebula environment, although the effects on the pulse shape are detrimental to our efforts to time the Crab Pulsar and hence our ability to study glitches and spin-evolution. 

The structure of the paper is as follows: we describe the effect of the nebula on radio pulses in Section \ref{CrabPaperDMandScatterIntro}. In Section \ref{CrabPaperObservations} and Section \ref{CrabPaperDMandScatteringDetermination}, we describe our observations and methods for measuring the pulse shape variations due to the nebula, using our 30-yr data set. We present the results in Section \ref{CrabPaperResults}, 
and apply our results to the timing data of the pulsar in Section \ref{CrabPaperTiming}. We discuss our findings and make general comments in Section \ref{CrabPaperConclusions}.

\begin{table*}
\caption{Summary of the backends used on the Lovell and the 42-ft telescopes for our data set.} 
\label{table:backends}
\centering
\begin{scriptsize}
\begin{tabular} {c c c c c c c c c} 
\hline \hline                        
Telescope & Filterbank & Centre Freq. (MHz) & Bandwidth (MHz) & $n_{\text{chan}}$ & $T_{\text{obs}}$ (typical) & Cadence (typical) & $n_{\text{bins}}$ & Usage \\ [0.5ex] 
\hline                  
Lovell & AFB & 1396 - 1416 & 64 & 64 & 12\,min & weekly & 400 & Sept.\,1982 - Apr.\,2010 \\ 
 & & & & & & & & (MJD\,45213 - 55287)\\ 
  & DFB & 1380 - 1532 & 384 & 768 & 12\,min & daily & 1024 & Jan.\,2009 - Mar.\,2014 \\
  & & & & & & & & (MJD\,54832 - 56739)\\ 
\hline \hline
42-ft & AFB & 607 - 612 & 4 & 32  & $10$\,hr & daily & 400 & Jan.\,1984 - Oct.\,2012 \\
& & &  & &  & & & (MJD\,45700 - 56201)\\
  & COBRA2 & 610 & 5,10 & 20,40 & $10$\,hr & daily & 1024 &Nov.\,2011 - Apr.\,2014 \\
 & & & & &  & & & (MJD\,55866 - 56756)\\
\hline 
\end{tabular}
\end{scriptsize}
\end{table*}
\begin{figure*}
\includegraphics[scale=0.41]{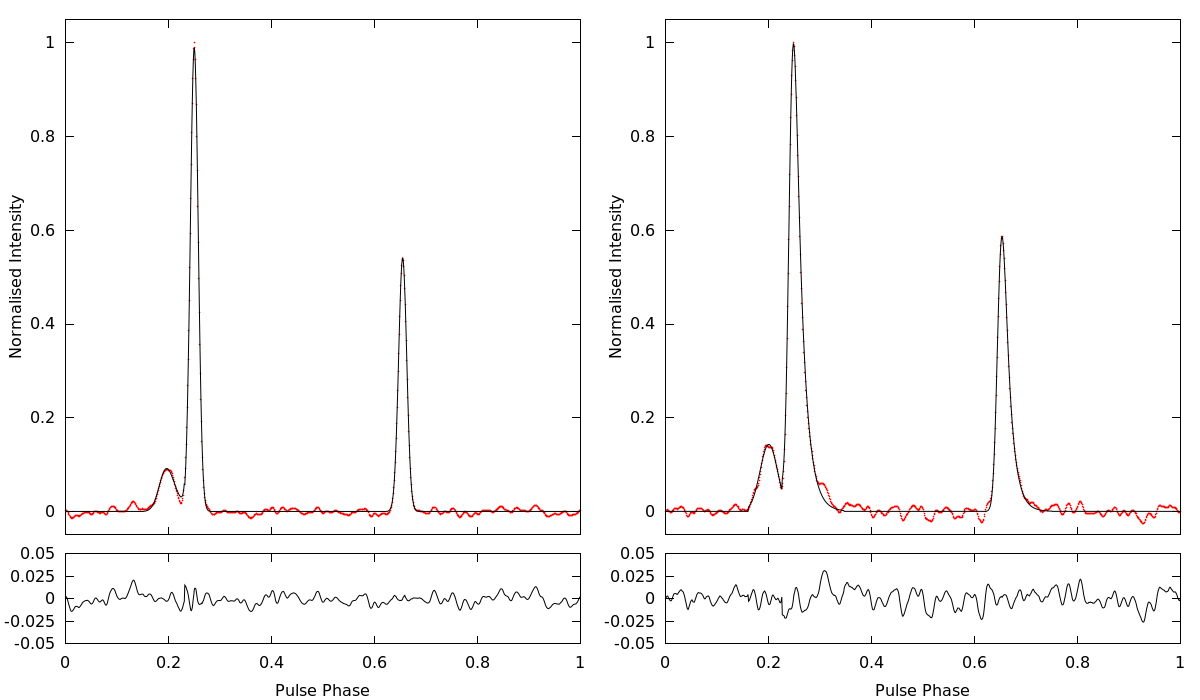}
	\centering
	\caption{Examples of the exponentially-modified Gaussian fit to the main pulse and interpulse components of the Crab Pulsar at 610\,MHz, and the residual between the best-fit model and the data for low and high scattering time scales. The AFB data were each averaged over 15 days of observation, centred on MJD\,53140 (left), and MJD\,48489 (right). The time scales of the scattering tails, calculated from the weighted mean of the MP and IP components, are $0.000 \pm 0.003$\,ms and $0.46 \pm 0.01$\,ms respectively. In both cases, the root mean square variation of the residual is consistent with the off-pulse noise of the profiles (0.006 and 0.005 respectively for MJD\,53140, 0.009 and 0.01 for MJD\,48489), indicating a close agreement between the model and the data. Note that fluence of the pulse components is conserved during scattering, which causes the peak amplitude of the (intrinsically broader) precursor to increase relative to the main pulse, due to the normalisation. Also note that the raw sub-integration data were folded into wide phase bins, which were then re-sampled with finer time resolution in the later analysis, prior to realignment and co-adding to form the integrated profiles, which causes the noise in the phase bins to appear highly correlated.} 
\label{fig:fit}
\end{figure*}

\section{Dispersion and Multipath Scattering} \label{CrabPaperDMandScatterIntro}
A cold plasma of free electrons affects radio propagation and hence the observed pulse shape of a pulsar in two dominant ways; through dispersion and through multipath scattering. Dispersion is related to the total number of electrons along the line of sight $l$ to the pulsar and is quantified by the dispersion measure $\text{DM}=\int n_{\text{e}}\,\mathrm{d}l$, where $n_{\text{e}}$ is the electron number density (see \citealp{lk05} for a review). 
Dispersion causes the group velocity of the signal to be frequency dependent, and delays the received pulse by an amount $\tau_{\text{DM}} \propto f^{-2}$, where $f$ is the observing frequency. If left uncorrected, the delay causes the received signal to sweep across the observing bandwidth, resulting in a broadening of the observed profile.
The radiation from the Crab Pulsar passes through three distinct regions of free electrons along the line of sight which may affect the pulse shape: the surrounding nebula, the ISM, and the Solar wind. For measured DM variations of $\Delta \text{DM}\!\sim\! 0.03\,\text{cm}^{-3}\,\text{pc}$ reported by \citeauthor{klj+08} \citeyearpar{klj+08}, and a nebula radius of 1.7\,pc \citep{hes08}, this corresponds to an average electron number density variation of $\Delta n_{\text{e}}\!\sim\!0.018\,\text{cm}^{-3}$. As a comparison, DM variations in typical pulsars at similar distances over short time scales have a magnitude of $\Delta \text{DM} \sim 10^{-4}$\,cm$^{-3}$\,pc \citep{kcs+13} which, for the Crab Pulsar at a distance of $\sim\!2200$\,pc (\citealp{tri73}, \citealp{kcg+08}), would correspond to $\Delta n_{\text{e}}\!\sim\!2 \times 10^{-6}$\,cm$^{-3}$. 
Therefore, in the case of the Crab Pulsar, it is likely that any observed changes in pulse shape are due to the influence of the nebula, since the electron number density variations within the nebula are several orders of magnitude higher than those in the ISM.

Scattering by the refractive index variations due to small-scale density fluctuations in the ISM causes incident rays to have scattering-angle-dependent path lengths, delaying them relative to unscattered rays. The delay experienced by scattered rays causes the observed pulse profile of a pulsar to be broadened, and to exhibit a characteristic exponential scattering tail.
The influence of scattering on pulse profiles is described in a series of papers by \citeauthor {wil72} (\citeyear{wil72}, \citeyear{wil73}, \citeyear{wil74}), for scenarios in which the scattering can be modelled by a thin screen, an extended screen, or by multiple screens, located along the line of sight between the source and the observer. 
The line of sight to the Crab Pulsar has two scattering screens; in the ISM and in the nebula, but as discussed above, electron density fluctuations in the nebula are much greater than in the ISM. In cases where scattering is dominated by one screen, \cite{wil73} noted that the multi-screen model result tends towards that of a simple single-screen model.
For this reason, we assume a thin screen when modelling the observed scattering in the Crab Pulsar. 
The magnitude of observed scattering scales with the placement of the scattering screen along the line of sight.  
As we are only interested in the effect on the observed pulse shape and the time scale over which relative pulse-shape variations occur, the placement of the screen is not relevant to our measurements, meaning we are free to use the simplest case: a thin scattering screen midway between the Earth and the pulsar, which has previously been shown to provide a good approximation of observed pulse shapes in this source (e.g. \citealp{cbh+04}). 

The broadening of the pulse profile due to scattering is usually modelled by convolving an unscattered pulse shape with a one-sided exponential.
If the intrinsic pulse can be modelled as a Gaussian, as is the case for the components of the Crab Pulsar's profile \citep{lgw+13}, it has been shown \citep{mck14} that an exponentially-modified Gaussian function provides a good approximation of the scattered pulse shape (see Appendix \ref{appendix}), and can be used to recover the parameters of the intrinsic Gaussian-like pulse. 
However, we note that so-called extreme scattering events occur occasionally, and are not adequately modelled using a single exponential scattering tail (e.g. \citealp{lt75}, \citealp{bwv00}, \citealp{lpg01}).
Additionally, as the exponentially-modified Gaussian function is defined over an infinite range, it can only be fit over a finite window if the exponential component is shorter than the window size. This would prevent the function from modelling scattered profiles where the length of the exponential component is greater than the pulse period. It also means that profiles with components very close to the edge of the phase window must be rotated in phase before fitting the function.

\begin{figure*}
	\includegraphics[scale=0.72]{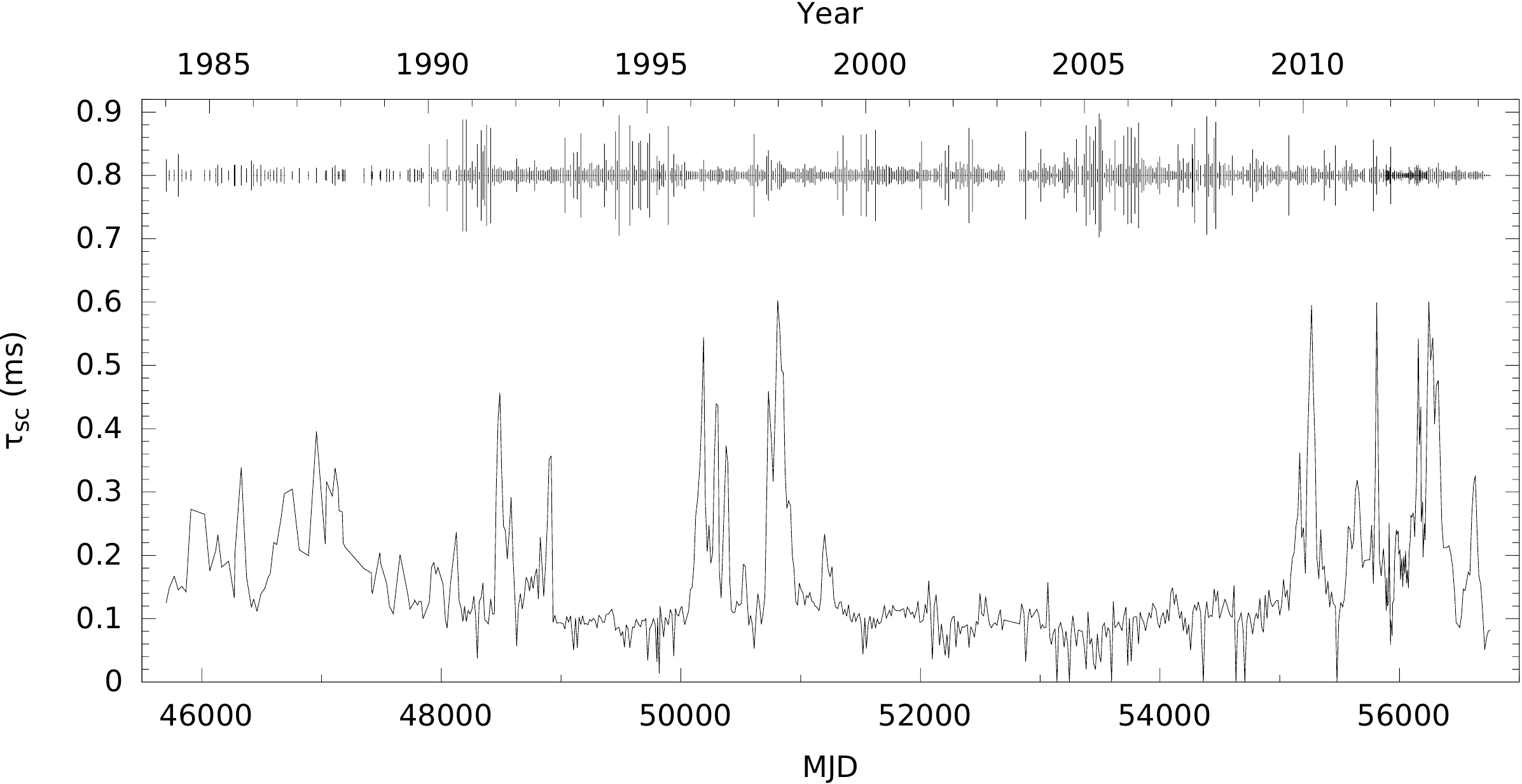}
	\centering
	\caption{Scattering time scales $\tau_{\text{sc}}$ 
in the Crab Pulsar at 610\,MHz between MJD\,45704 and 56756. Values were measured by fitting the exponentially-modified Gaussian function to the main pulse and interpulse of 610-MHz profiles averaged over 15 days, and taking $\tau_{\text{sc}}$ to be the weighted mean of the two measurements. Periods of high scattering are easily identifiable against the 
median value $\sim 0.13$\,ms. Error bars for the data points are displayed above to preserve clarity.}
\label{fig:scatter}
\end{figure*}
\begin{figure}
\includegraphics[scale=0.7]{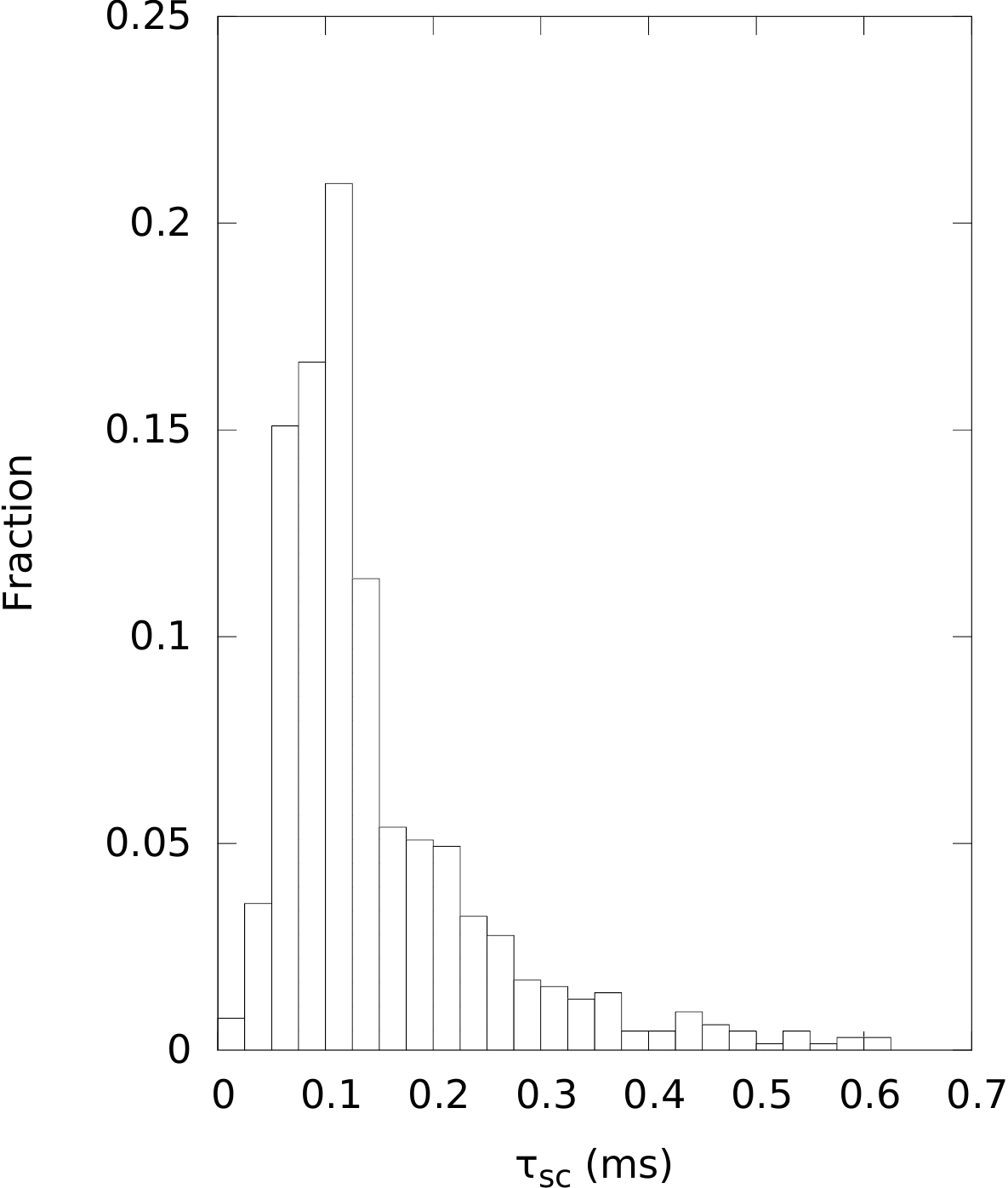}
	\centering
	\caption{Distribution of measured scattering time scales at 610\,MHz, using the data presented in Figure \ref{fig:scatter}. Approximately $64\%$ of the distribution lies in the range ${0.05\,\text{ms}<\tau_{\text{sc}}<0.15}$\,ms, where a clear peak at the time scale $\tau_{\text{sc}}\!\sim\!0.13$\,ms is visible, followed by a decay to rarer instances of high scattering.}
\label{fig:dist}
\end{figure}
\begin{figure*}
\includegraphics[scale=0.725]{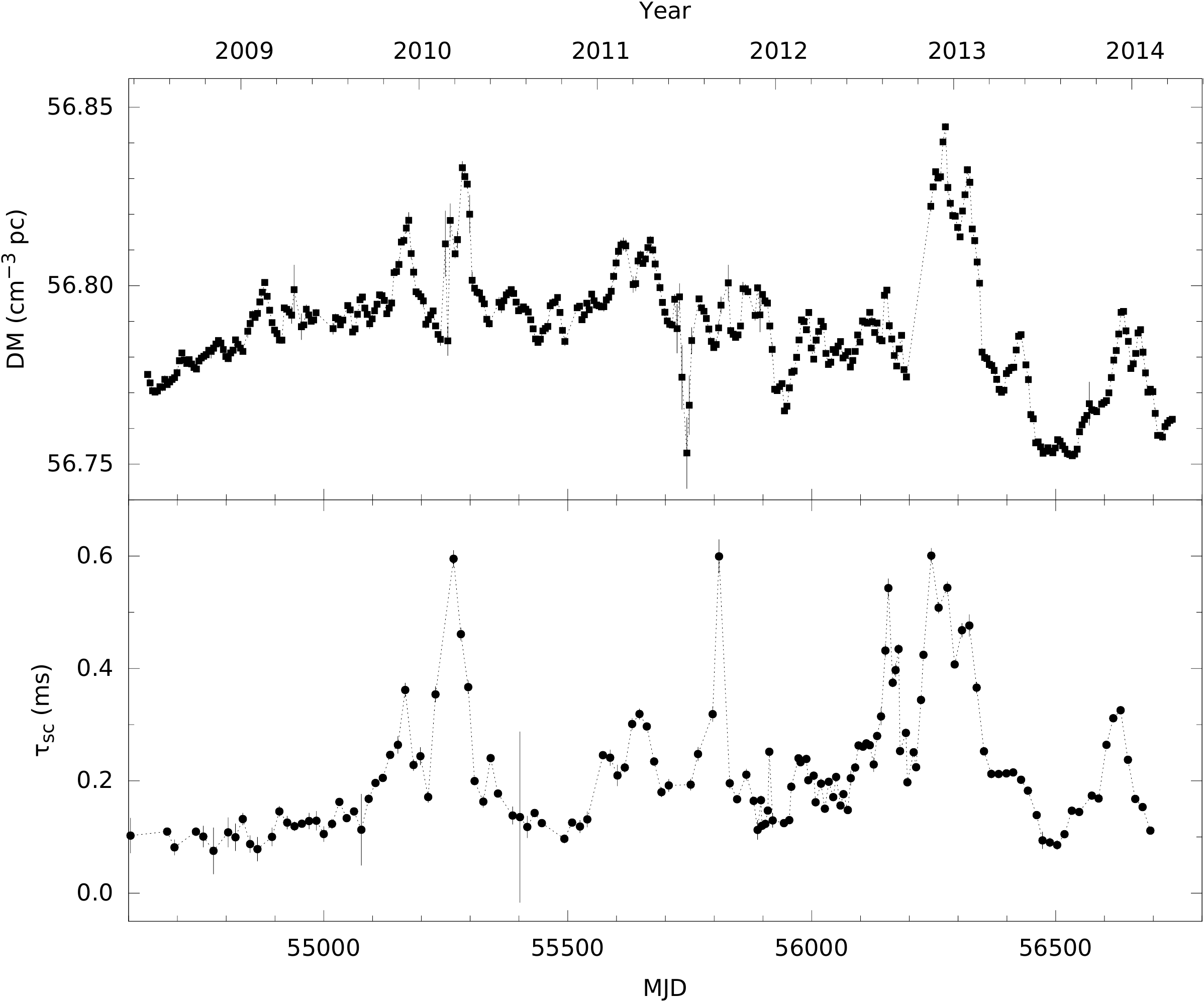}
	\centering
	\caption{Measured scattering time scales $\tau_{\text{sc}}$ at 610\,MHz (below, circles), and DM measurements over the period MJD\,54639 to 56739 (above, squares). The $\tau_{\text{sc}}$ measurements have been used to remove the excess DM arising from the extra TOA delays due to scattering. Sharp increases and decreases in scattering correspond to equivalent changes in DM over this period, which are closely correlated.}
\label{fig:dm_tau}
\end{figure*}
\section{Observations} \label{CrabPaperObservations}
The data used in this study were obtained at Jodrell Bank Observatory using almost daily observations over a 30-year period. The majority were recorded using the 42-ft telescope observing at frequencies near 610\,MHz, typically with 10 hour integrations (see Table \ref{table:backends} for details). These data are supplemented with observations around 1400\,MHz using the 76-m Lovell Telescope, with a typical integration time of 12 minutes. 

The Lovell Telescope receiver is cryogenically cooled and detects both circular polarisations. The received signals were dedispersed incoherently using either an analogue filterbank (AFB) or a digital filterbank (DFB). The 42-ft telescope uses a room temperature receiver and also detects both circular polarisations, which are dedispersed using either an AFB or the COBRA2 backend. In all cases, data from both polarisations were filtered, digitised, and dedispersed (summarised in Table \ref{table:backends}). More in-depth information on the operation of these backends can be found in \cite{sl96} (AFB) and \cite{mhb+13} (DFB). Observations using the COBRA2 backend were dedispersed coherently using \textsc{DSPSR} \citep{vb11}, and those using the AFB and DFB were dedispersed incoherently. The frequencies and uncalibrated polarisations of each of the dedispersed observations were summed, and the resulting total-intensity pulses were folded, producing average pulse
profiles, from which times of arrival (TOAs) were derived using a standard template corresponding to the average profile shape at the observing frequency (more details below).

Due to the presence of glitches, timing noise, and DM variations observed in the Crab Pulsar, the folding used a variety of ephemerides relevant to the period in which the observation took place, with each ephemeris accounting for a period of around 200 days. Data were folded online using these ephemerides, which were all sufficiently accurate as to cause no significant pulse-shape broadening due to DM errors or rotational variations. We note that the 4-10\,MHz bandwidth used in our 610\,MHz observations is small, and therefore small errors in the DM used for folding do not significantly alter the width of the folded pulse profile. 
Average profiles were created from the 610-MHz observations over 15-day periods, improving the S/N of the data, while preserving all but the shortest time-scale variations in the scattering properties of the Crab Pulsar. This was done by aligning profiles in phase, according to the timing ephemerides, and co-adding the profiles. This yielded a total of 649 average profiles, at time resolutions of approximately \SI{33}{\micro\second}.

\section{Determination of DM and Scattering} \label{CrabPaperDMandScatteringDetermination}
The 15-day average profiles were each normalised so that the peak intensity of the main pulse (MP) was always equal to unity, and rotated so that the peak of the MP was centred at a phase of 0.25. The exponentially-modified Gaussian function was then fit separately to the MP, interpulse (IP), and precursor components of the average profiles. Although scattering observed in the Crab Pulsar is generally quasi-exponential, extreme scattering events have been observed to deviate from an exponential (\citealp{lt75}, \citealp{bwv00}, \citealp{lpg01}). This means that an exponentially-modified Gaussian function does not necessarily model the true scattering function. Such events represented $\sim\!17\%$ of our initial data set, and were excluded from our analysis. The exponential scattering model was found to be a good fit to the non-anomalous scattering, with the residuals between the model and the fit being consistent with the off-pulse noise. The ratio of the measured scattering time scales of both components was found to agree well, with a value $\tau_{\text{IP}}/\tau_{\text{MP}}=1.00\pm0.18$ (where here and elsewhere, the quoted errors are $1\sigma$) for the whole data set.
The parameters for the fits to the MP and IP were combined, and weighted by the relative S/N, to determine the average scattering and DM parameters at each epoch. Due to the low S/N of the precursor component, this was not included in the calculation of the average scattering. As errors in the DM used for folding the average profile can broaden the pulse shape, and therefore have a covariance with pulse-shape variations due to scattering, we tested our method for measuring the scattering time scales, by creating simulated profiles with a known scattering time scales, and varying folding DM errors. The simulated profiles were created using the \textsc{PSRSALSA}\footnote{http://www.jb.man.ac.uk/~pulsar/Resources/psrsalsa.html} suite \citep{wel16}, with DM errors in the range $0.001\,\text{cm}^{-3}\,\text{pc}<\sigma_{\rm DM}<0.5\,\text{cm}^{-3}\,\text{pc}$, and the scattering time scales were measured using the same method we describe above. Only small errors in measured scattering were introduced ($<10\%$ $\tau_{\text{sc}}$), as the DM error increased, and were always negligible compared to the measurement precision. We are therefore confident that our scattering measurements are not appreciably contaminated by small errors in the folding DM.

Values for the DM were derived from the TOAs during the period MJD\,54639 to 56739, where regular observations at both 610\,MHz and 1400\,MHz were taken. We define `stride fitting' as a technique where a portion of the data is fit for over a window of constant length, before the window is shifted by a constant value and fitting is repeated. The TOAs were analysed with the \textsc{PSRTIME}\footnote{http://www.jb.man.ac.uk/pulsar/observing/progs/psrtime.html} software package, and DMs were determined using a stride fit over 10-day windows with a 5-day long stride, and an ephemeris with an initial DM of 56.78\,$\text{cm}^{-3}$\,pc. The stride fit was performed for DM, period, and period derivative, using TOA weights which were proportional to the inverse square of their uncertainty.
As the DM value fluctuates over a range of $\sim\!0.1$\,$\text{cm}^{-3}$\,pc, only DMs with
uncertainty less than 0.01\,$\text{cm}^{-3}$\,pc were included in the analysis, which allowed small-scale variations to be distinguished. The TOAs were calculated using a standard fixed template. However, as described above, scattering can shift the centroid position and therefore introduce an apparent delay, which can mimic an additional DM term. This effect was mitigated by calculating the excess DM due to the additional scattering delay 
by measuring the centroid shift of the scattered profile relative to the standard template, converting the delay to a DM value, and subtracting it from the initial DM measurement. Pulse broadening due to differences between the folding DM and true DM were not considered, as these are much less significant than the broadening due to scattering (a few microseconds compared to a hundreds of microseconds).
The corrections applied to the measured DMs have a maximum size of 0.03\,cm$^{-3}$\,pc i.e. smaller than the measured DM variations, but still 
large compared to the measurement uncertainty. Although the AFB observations were dedispersed incoherently, the DM smearing at 610\,MHz for typical DM values is $\sim1$ phase bin ($\sim\SI{33}{\micro\second}$), and therefore does not significantly contaminate our measurements of the scattering time scales.

It has recently been shown that DM measurements made at widely-separated frequencies may be incompatible, due to scattered pulses sampling different path lengths through the ISM, which gives rise to so-called `chromatic DM' effects \citep{css16}. Using the equations presented in Section 2.3 of \cite{css16}, we estimate the maximum contribution of this effect to be $\sim 0.001$\,cm$^{-3}$\,pc, which is in general lower than our measurement precision and the size of the variations we observe.

\begin{figure*}
	\includegraphics[scale=0.72]{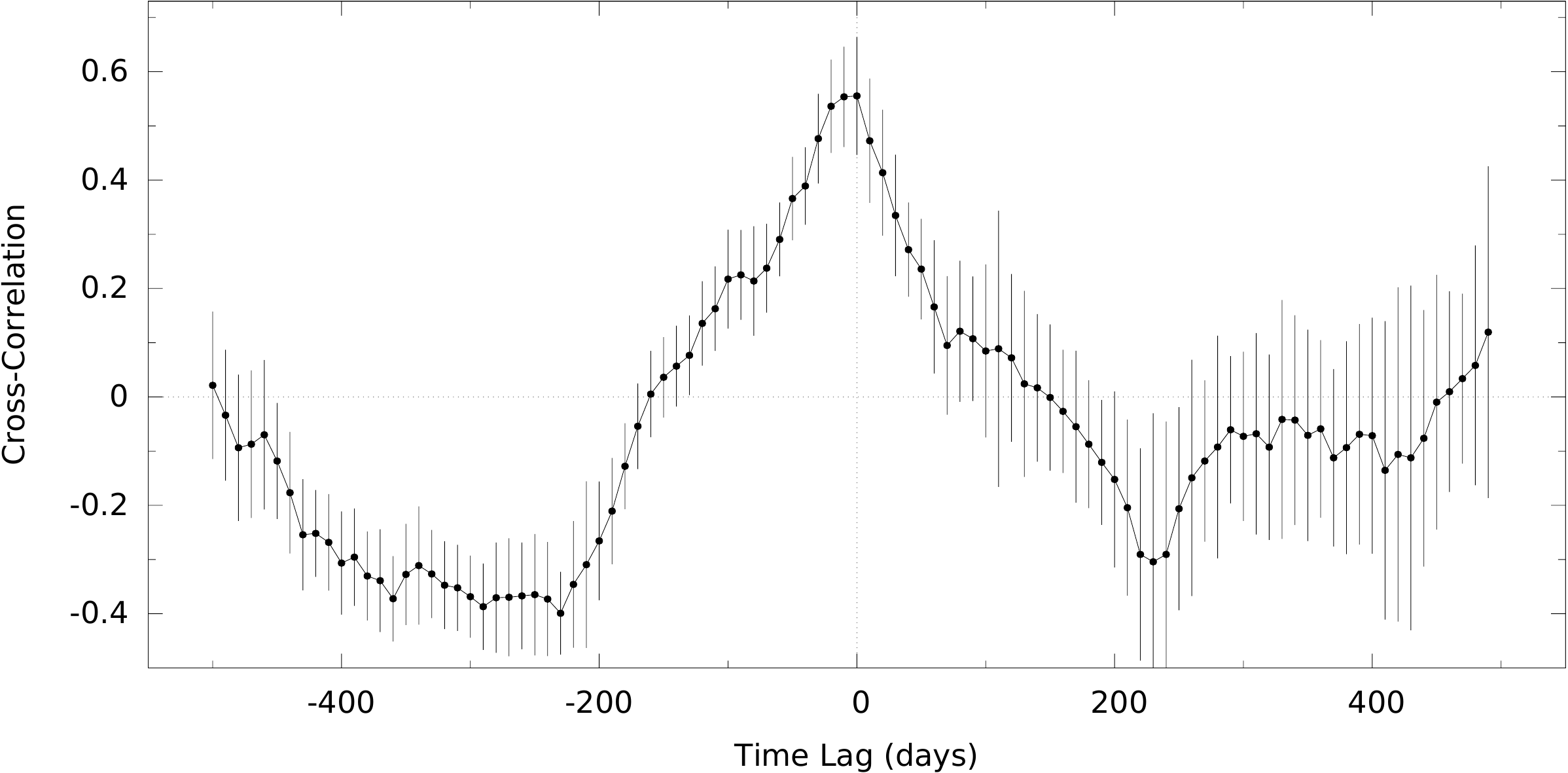}
	\centering
	\caption{The cross-correlation function of the measured changes in $\tau_{\text{sc}}$ and DM in the period MJD\,54639 to 56739, calculated using the data presented in Figure \ref{fig:dm_tau}. Uncertainties were quantified by varying the input values of the DM and scattering data sets, by drawing values for each bin from Gaussian distributions defined by the measurement and the error bar of the input values, and repeating the cross-correlation measurement for 1000 trials. The plotted values and uncertainties of each data point are the average and $1\sigma$ range of all of these trials, and include the statistical error due to the sample size of each lag bin. The peak at zero lag indicates that there is no significant delay between these two data sets, and that the DM and $\tau_{\text{sc}}$ variations are highly correlated.}
\label{fig:crosscor}
\end{figure*}

\section{Results} \label{CrabPaperResults}
\subsection{Scattering and DM Variations}
Scattering time scales were measured for each of the 15-day folded 610-MHz pulse profiles (Figure \ref{fig:scatter}). Plotting the variation in scattering with time clearly shows periods of high scattering, with rare instances of extremely low
scattering.
The range 0.05\,$\text{ms}<\tau_{\text{sc}}<0.15$\,ms accounts for $\sim\!64\%$ of the measured scattering values, with the median value $\tau_{\text{sc}}\!\sim\!0.13$\,ms representing a `quiet' baseline, about which variations take place. Figure \ref{fig:dist} shows a histogram of the measured scattering time scales, where 0.13\,ms is clearly the preferred time scale, and longer time scales are much more rare. This allows us to define the values outside this range as instances of low/high scattering. 

\begin{figure*}
	\includegraphics[scale=0.72]{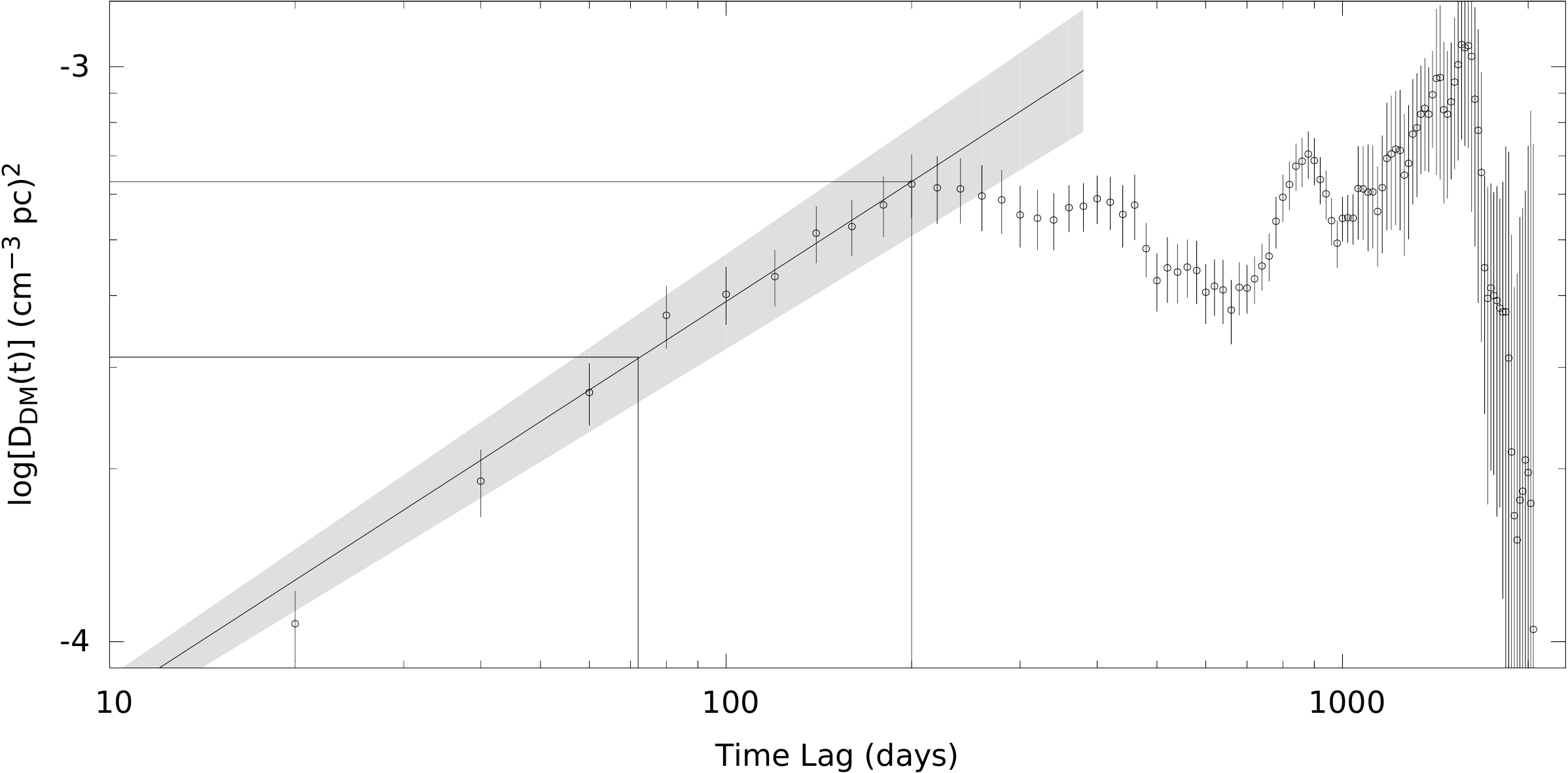}
	\centering
	\caption{Structure function computed for the DM data set presented in Figure \ref{fig:dm_tau}, with the saturation value and characteristic time scale highlighted. The power law fit to the structure regime is plotted, with the grey region representing the $1\sigma$ uncertainty. Uncertainties in the data points were quantified using the same approach as outlined in Figure \ref{fig:crosscor}. The structure function appears to saturate at a time lag of $\sim\!200$ days, which allows a characteristic time scale of $\sim70$ days to be estimated from the power law fit to the structure regime. The characteristic time scale corresponds to density variations on length scales of $\sim\!6.4$\,a.u. ($\sim2.4\times10^{-5}$\,pc).}
\label{fig:structure}
\end{figure*}

The relationship between scattering measured at 610\,MHz and DM was investigated over the period MJD\,54639 to 56739, where there were sufficient multifrequency observations for high-precision measurements of the DM to be made (Figure \ref{fig:dm_tau}). 
The scattering time scales measured in this period are notably high compared to the previous $\sim\!3000$ days, with $\tau_{\text{sc}}$ routinely exceeding 0.3\,ms. 
Inspection of Figure \ref{fig:dm_tau} indicates a correlation between $\tau_{\text{sc}}$ and DM; a relationship previously observed in data from the Crab Pulsar by \cite{klj+08}, and for which evidence has been presented using observations of canonical pulsars (e.g. \citealp{kcw+18}) and millisecond pulsars (e.g. \citealp{cks+15}). This was confirmed by a cross-correlation analysis of the DM and $\tau_{\text{sc}}$ data, which revealed 
a non-significant lag of $-10\pm10$\,days between the two data sets, with a correlation coefficient of $0.56\pm0.01$ (both before and after removing the scattering contribution to the DM measurements), indicating a close correlation between DM and scattering (Figure \ref{fig:crosscor}). The measured lag is consistent with the 30-day delay between scattering (at 111\,MHz) and DM peaks reported by \citeauthor{klj+08} (\citeyear{klj+08}), but also consistent with zero. As can be seen in Figure \ref{fig:scatter}, no long-term secular variation in scattering time scales are apparent over the 30-yr period of our data set.

\begin{figure*}
	\includegraphics[scale=0.693]{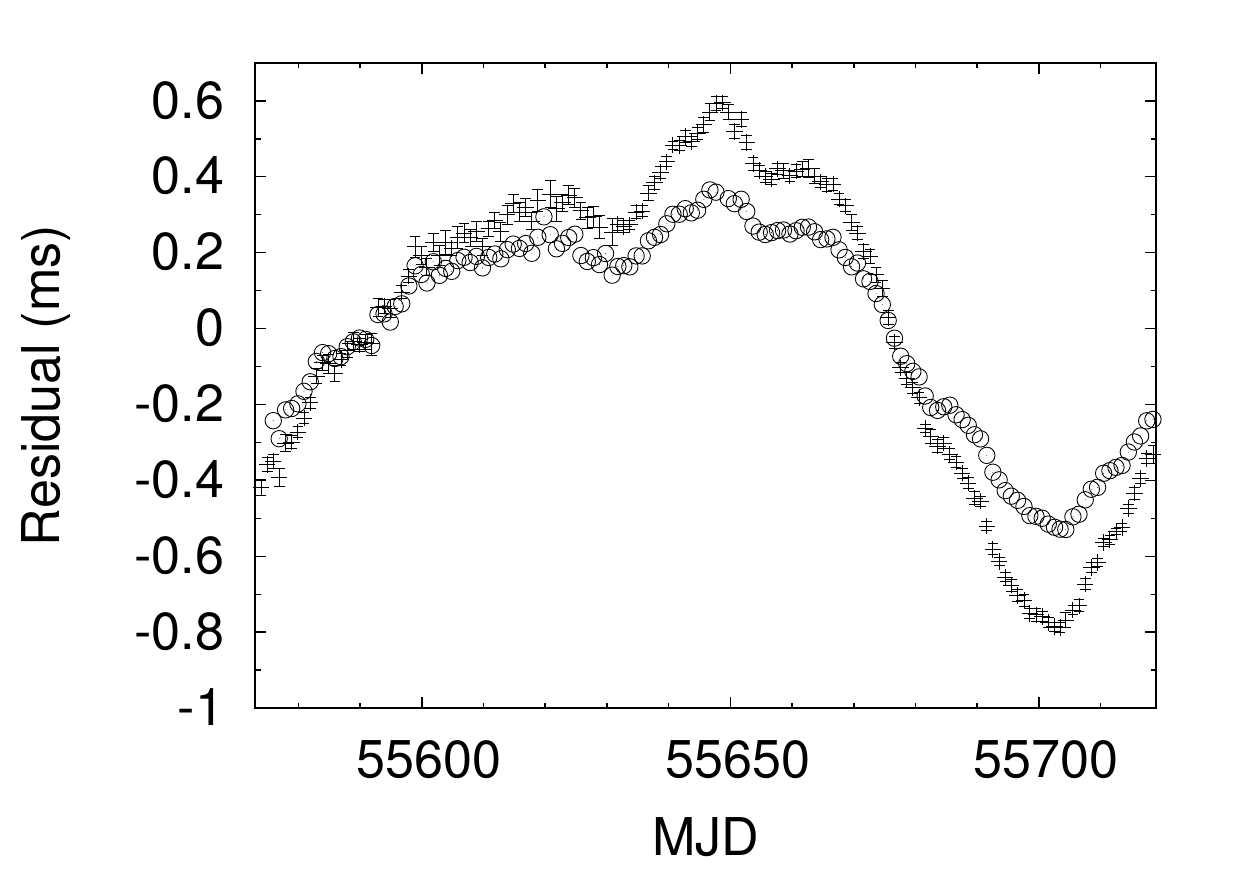}
	\includegraphics[scale=0.693]{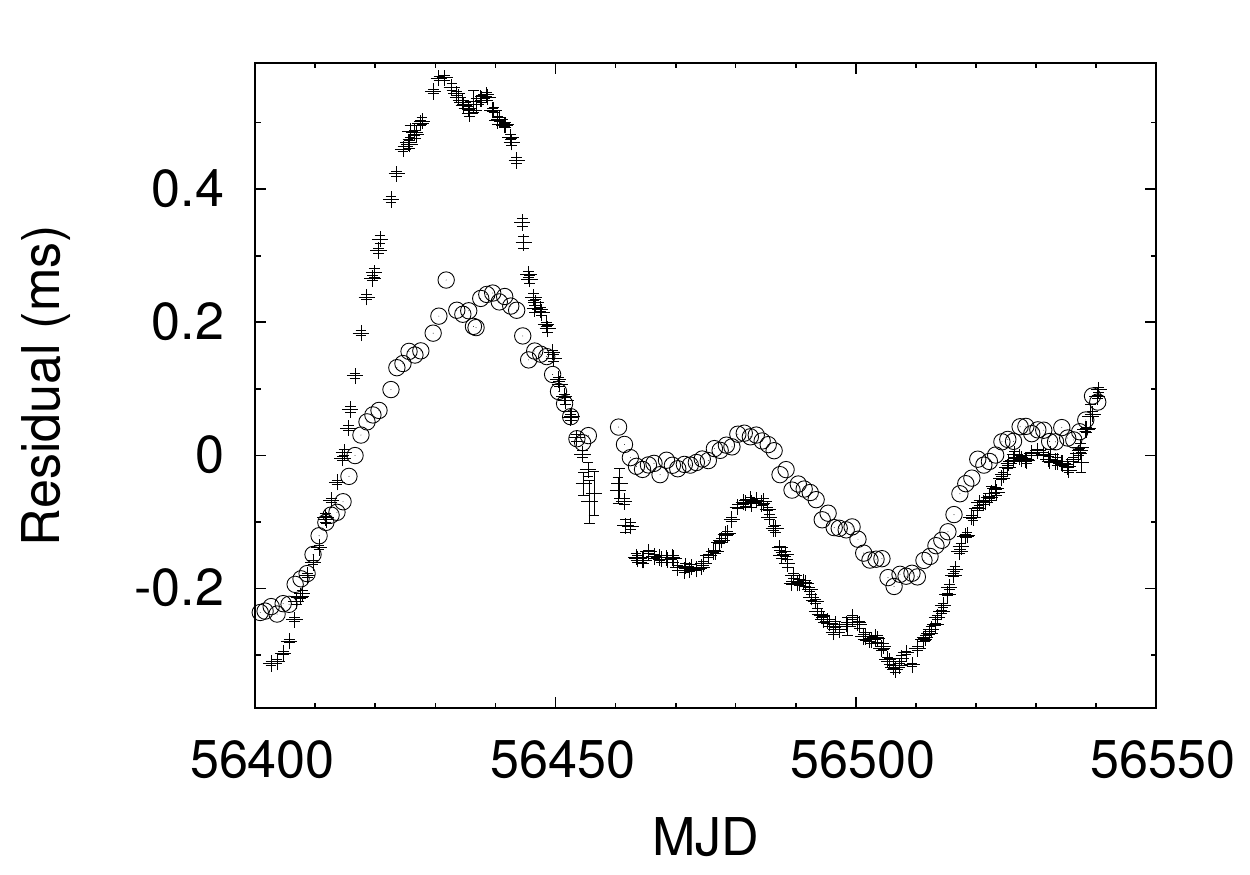}
	\centering
	\caption{Examples of timing residuals derived from TOAs obtained using standard templates ($+$ symbols), and templates corrected for the measured DM and scattering (open circles), in the period MJD\,55573 to 55719 (left) and MJD\,56400 to 56550 (right) for observations using the 42-ft at 610\,MHz. Both sets of TOAs used the same ephemeris, relevant to the epoch. In both epochs, accounting for the variations in scattering when deriving TOAs, and DM in the timing analysis resulted in lower RMS residuals (left: $\text{RMS}=\SI{403}{\micro\second}$ improved to $\text{RMS}=\SI{261}{\micro\second}$; right: $\text{RMS}=\SI{256}{\micro\second}$, improved to $\text{RMS}=\SI{124}{\micro\second}$).}
\label{fig:residuals}
\end{figure*}

\subsection{DM Structure Function Analysis}
A useful statistical tool for quantifying turbulence in time series is the structure function, which describes the mean-squared variance of the time series for pairs of measurements separated by a time lag $t$. It has recently been applied to studies of ISM turbulence (e.g. \citealp{lcc+16}). The structure function is given by
\begin{equation}
D_{\text{DM}}(t_{k})=\sum_{j=1}^{N} \frac{(\text{DM}_{j}-\text{DM}_{j+k})^{2}}{W_{k}},
\end{equation}
where $W_{k}$ is the number of data pairs separated by a time lag $t_{k}$. The structure function of a turbulent medium has three distinct regimes: the \textit{noise regime} at time lags shorter than the shortest time scale of the underlying process, the \textit{structure regime} where a power law relationship is obeyed (index $\sim\!5/3$ in the case of Kolmogorov turbulence), and the \textit{saturation regime} at long time lags, where the structure function plateaus or begins to oscillate. The characteristic time scale of the variations is defined as time lag at which the value of the structure function is half the saturation value.

The structure function was computed for the DM time series in Figure \ref{fig:dm_tau}, using a lag of 20 days as a conservative lower limit, to ensure that the noise regime would be avoided (Figure \ref{fig:structure}). The structure regime and saturation regimes are both clearly visible. The power law index of the structure regime is $0.69\pm0.04$, much less than the Kolmogorov value of $\sim\!5/3$. The non-Kolmogorov behaviour of the data set is not unexpected, if variations are not stochastic, and are instead caused by discrete regions of high-density, compared to the bulk nebula. Computing the DM structure function for the first 260 days of the data set (where there are no large DM `events') revealed a power law relationship with an index of $1.85\pm0.09$ for the structure regime, which is in much closer agreement with the prediction for Kolmogorov turbulence.

The saturation value was estimated by taking the derivative of the function and identifying where the gradient changes. The gradient turns over at a lag of $\sim\!200$ days, which allowed the characteristic time scale to be estimated as $\sim\!70$ days. This preferred time scale corresponds to the typical length of the DM `events' seen in the time series, which was confirmed by computing the structure function of simulated DM time series exhibiting similar DM events, using a model where DM variations are drawn from a normal distribution defined by the magnitude of variations we observe, and injecting larger variations of similar size to those observed in our data set. Assuming that the variations we see are due to the motion of the pulsar through the nebula providing different lines of sight, and taking the transverse velocity of the pulsar to be 120\,km\,s$^{-1}$ \citep{kcg+08}, a typical length scale of $\sim\!6$\,a.u. ($\sim\!2\times10^{-5}$\,pc) was calculated for the high-density regions of the nebula.  
We note that the sharp wisps seen in optical observations of the Crab Nebula have lateral scales of $\sim\!1.6\times10^{-3}$\,pc \citep{hss+95}, corresponding to a 13-yr time scale. As features in the 30-year scattering data on time scales comparable to this are not obvious, and variations instead are dominated by those at much shorter time scales, this implies that, if any wisps have crossed the line of sight to the pulsar in this time, they must be of a much lower density than the structures responsible for the sharp increases that we observe.
There is a hint of a long-term quadratic trend in the 6-yr DM data set presented in Figure \ref{fig:dm_tau}, which would be in closer agreement with the time scales seen in optical observations, a longer data set would be required to rule out stochastic variations as the cause.
The time scales we measure are instead consistent with the estimated scale of structures previously associated with scattering and dispersive events in the Crab Nebula (\citealp{glj11}, \citealp{klj+08}). We interpret this as evidence for discrete structures in the nebula smaller than those resolvable in the optical as being the cause of the large increases in DM and scattering we observe.

\section{Precision Timing of the Crab Pulsar} \label{CrabPaperTiming}
Although the Crab Pulsar has been observed for decades, the precision of the timing measurements has been restricted by the presence of timing noise and 
ISM effects. The scattering and DM measurements we have made can be used to improve our ephemerides and pulse profiles, allowing us to better isolate the astrophysical effects of the timing noise.
For example, the 
pulse profile at 1400\,MHz and 610\,MHz has been shown to evolve in both frequency and time \citep{lgw+13}, with the relative flux density of the IP to the MP linearly decreasing at a rate of $\sim\!0.172 \ \text{century}^{-1}$ at 610\,MHz and $\sim\!0.17 \ \text{century}^{-1}$ at 1400\,MHz. 
Measurements at 1400\,MHz and 610\,MHz have also shown that the separation between these two components is increasing at a rate of $\sim\!0.5^{\circ} \ \text{century}^{-1}$, making constant standard templates unsuitable for long-term precision timing.
In addition, 26 timing glitches have been observed in the Crab Pulsar's rotation throughout its observational history (two most recent are reported in \citealp{slb+17} and \citealp{ejb+11}, and see \citealp{els+11} for more information). 
These are sudden increases in the pulsar's spin frequency and spin-down rate, followed by a short-term transient recovery, with fractional rotation frequency changes in the Crab Pulsar ranging from $\Delta \nu/\nu=0.8\times10^{-9} $ to $214\times10^{-9} $ \citep{eas+14}.
The combined effect of these glitches has made it useful to use a catalogue of ephemerides for the Crab Pulsar timing solution (The Crab Pulsar Monthly Ephemeris\footnote{http://www.jb.man.ac.uk/pulsar/crab.html}, \citealp{lpg93}), as opposed to a single set of rotational parameters. By accounting for timing glitches, it has been possible to accurately measure the braking index of the Crab Pulsar \citep{ljg+15}. The measurements we have made of the scattering and DM variations in the Crab Pulsar can potentially be used to improve our ephemerides, and better isolate astrophysical effects.
Additionally, the Crab Pulsar exhibits significant timing noise, which \cite{lhk+10} demonstrated can be correlated with pulse shape changes, in 6 out of the 17 pulsars in their study. As ISM effects can be a source of pulse shape variations, and potentially hinder the measurement of any intrinsic changes on shorter time scales, disentangling the time-variable DM and scattering influence on the radio pulse shape would increase our precision in studies of timing noise in the Crab Pulsar.

Techniques for removing the influence of scattering 
on pulsar timing are not widely used at present, although the possibility is being explored for use with millisecond pulsars included in pulsar timing arrays (PTAs), where minimising the noise of the timing data is necessary to improve the precision of gravitational wave detection experiments.
Most notably, cyclic spectroscopic analysis of PTA pulsars has been shown to be effective in measuring and correcting some of this contamination by scattering of the pulse shape \citep{dem11}, but this is not feasible for use with historical data. Attempts at mitigating the statistical error of DM from timing observations have included modelling it as an additional stochastic parameter in timing noise analysis e.g. the \textsc{TempoNest}\footnote{https://github.com/LindleyLentati/TempoNest/} package (\citealp{lah+14}), estimation via interpolation using polynomials (e.g. \citealp{fwe+12}) or piecewise linear functions \citep{kcs+13}, and through sub-band-dependent pulse shape variations \citep{lkd+17}.

We have decided to use the DM and scattering time scale measurements from the previous sections to construct custom templates for periods in which high scattering had been observed, as a means of testing the feasibility of correcting data using these measurements. 
The templates were constructed using the observed pulse shapes of the MP and IP components of the profile at 610\,MHz. 
These custom templates were used to re-derive the TOAs in each period using the standard template-matching procedure, and the resulting corrected TOAs were plotted alongside the original TOAs, with the same ephemeris used for both sets (Figure \ref{fig:residuals}).
The corrected TOAs were found to have lower RMS residuals, with the TOAs in the window MJD\,55573 to 55719 improving from $\text{RMS}=\SI{403}{\micro\second}$, to $\text{RMS}=\SI{261}{\micro\second}$, and those in the window MJD\,56400 to 56540 being improved from $\text{RMS}=\SI{256}{\micro\second}$ to $\text{RMS}=\SI{124}{\micro\second}$, when comparing the RMS of the residuals after correcting for DM and scattering variations, to the RMS of the original (uncorrected) residuals.

\section{Conclusions} \label{CrabPaperConclusions}
We have made measurements of the scattering variations in the Crab Pulsar over a 30-year period. The effects of any chromaticity in DM due to scattering are shown to be negligible. We have described a method for disentangling the effect of scattering from the profile of the Crab Pulsar, and used the scatter-corrected timing data to precisely measure DM variations in the Crab Pulsar timing data. Due to the close correlation between scattering and DM variations, and the rapid and large size of the variations when compared to those of other pulsars, we interpret these variations as being due to discrete structures within the nebula. We note that the apparent lack of features at longer time scales indicates that the contribution of the wisps seen in the optical is much smaller, and implies these structures are of a significantly lower density than the structures we can probe in our data set. We describe a method for removing the dispersive and scattering effects from the timing data, which we test by re-deriving TOAs during two periods of highly-variable scattering. In both cases, the RMS of the timing residuals was improved, but some noise remained, which indicates that not all of the observed timing noise in the Crab Pulsar is related to the nebula.

\section*{Acknowledgements}
JWM dedicates the work in this paper to his father, William McKee. We thank Francis Graham-Smith, Gemma Janssen, Benjamin Shaw, Patrick Weltevrede, and Cristina Ilie for very helpful discussions, and the anonymous referees for useful comments. The authors acknowledge the support of their colleagues at Jodrell Bank Observatory. This work was supported by the UK Science and Technology Research Council, under grant number ST/L000768/1.

\bibliographystyle{mnras}
\bibliography{psrrefs}{}
\bsp	
\label{lastpage}

\appendix
\section{Pulse Shape Changes Due to Multipath Scattering} \label{appendix}
For a thin scattering screen, rays from an intrinsically narrow pulse are scattered at random angles, with an RMS scattering angle $\theta_{0}$. From \cite{wil72}, the size of this scattering angle is related to the length of the exponential scattering tail by
\begin{equation}
\tau_{\text{sc}}=\frac{\theta_{0}^{2} D'}{2c},
\end{equation}
where $D'$ is the scattered ray path length, and $c$ is the speed of light. The mean square scatter per unit length $L_{\text{scatter}}$ is proportional to the RMS electron density fluctuations of the ISM $\Delta n_{\text{e}}$, and is given by {$L_{\text{scatter}}\!\sim\!r_{0}^{2}(\Delta n_{\text{e}})^{2}f^{-4}a^{-1}$}, where $r_{0}$ is the classical electron charge radius, and $a$ is the characteristic scale size over which the electron density varies. Using the small angle approximation, the average distance travelled by the scattered rays is related to the distance of the scattering screen from the source by
\begin{equation}
D'=\frac{D(D-\Delta)}{\Delta} ,
\end{equation}
where $D$ is the distance to the source, and $\Delta$ is the distance to the scattering screen. If we assume Kolmogorov turbulence, the size of the scattering tail scales with frequency as $\tau_{\text{sc}} \propto f^{-4.4}$.

The broadening of the scattered pulse is commonly modelled mathematically by the convolution of a function representing the intrinsic pulse with an exponential function $G(t,\tau)$. In the simple case where the intrinsic pulse is close to a Gaussian $F(t, A, \mu,\sigma)$, with amplitude, position, and width $A$, $\mu$ and $\sigma$, the scattered pulse shape is then
\begin{equation}
F(t, A, \mu,\sigma)*G(t,\tau)=H(t; A, \mu,\sigma,\tau)  .
\end{equation}
This convolution can be expressed analytically by an exponentially-modified Gaussian function \citep{gru72}, given by
\begin{equation}
H(t; A, \mu,\sigma,\tau)=A\exp\left\lbrace\frac{-(\mu-t)^{2}}{2\sigma^{2}}\right\rbrace\frac{\sigma}{\tau}\sqrt{\frac{\pi}{2}}
\text{erfcx}\left\lbrace\frac{1}{\sqrt{2}}\left(\frac{\mu-t}{\sigma}+\frac{\sigma}{\tau}\right)\right\rbrace , 
\end{equation}
where the scaled complementary error function $\text{erfcx}(t)$ and complementary error function $\text{erfc}(t)$ are
\begin{equation}
\text{erfcx}(t)=e^{t^{2}} \ \text{erfc}(t)  ,
\end{equation}
\begin{equation}
\text{erfc}(t)=1-\text{erf}(t)=\frac{2}{\sqrt{\pi}}\int_{t}^{\infty}e^{-x^{2}}\mathrm{d}x  .
\end{equation}
\end{document}